\magnification1200

\rightline{KCL-MTH-06-16}
\rightline{hep-th/yymmnnn}

\vskip 2cm
\centerline
{\bf   $E_{11}$ and Higher Spin Theories }
\vskip 1cm
\centerline{ Peter West}
\centerline{Department of Mathematics}
\centerline{King's College, London WC2R 2LS, UK}
\vskip 2cm
\leftline{\sl Abstract}
We consider the previous proposal of the author to use an extension
of spacetime obtained by taking the non-linear realisation of the
semi-direct product of $E_{11}$ with a set of generators belonging to one
of the fundamental representations of $E_{11}$. We determine, at low
levels, the symmetries that the associated  point particle  moving in
this generalised spacetime should possess and write down the
corresponding action. Quantisation of similar actions has been shown to
lead to the unfolded formulation of higher spin theories and we argue
that the generalised coordinates will lead in the non-linear realisation
to an infinite number of propagating higher spin fields.

\vskip2cm
\noindent

\vskip .5cm

\vfill
\eject
By studying  the properties of  the maximal supergravity theories in ten
and eleven dimensions it has been  conjectured [1] that a Kac-Moody
algebra of rank eleven, which was called $E_{11}$,  is a symmetry of M
theory. In this paper we will address some of the open questions which
arose in this work and show  that their resolution, at low levels,  
makes contact with the  higher spins theories that have  already been
studied in the literature [2-14].  As such, it will be useful  to explain
how the conjecture of reference [1] came about and what points were
unresolved.  In an earlier  paper [15] it was shown that  the formulation
of bosonic sector of eleven dimensional supergravity which included a 
 six form  as well as   three form gauge fields is  a non-linear
realisation of an algebra  $G_{11}$, together with the space-time
translations $P_a$, with the local sub-algebra being the Lorentz algebra.
The algebra $G_{11}$ consisted of 
$GL(11)$, which was responsible for gravity,  as well as  rank three and
rank six generators.  To be more precise reference [15]  took the
simultaneous non-linear realisation of $G_{11}$ and the space-time
translation with the conformal algebra and  found that the result was  
precisely  the equations of motion of eleven dimensional supergravity, up
to the specification of one constant.   
\par
The algebra $G_{11}$  contained the algebra $A_{10}$ and the Borel
subalgebra of
$E_7$  which was  therefore seen to be  a symmetry of eleven dimensional
supergravity [15].  However, $G_{11}$  looked incomplete in a number of
ways,  one of which was that it did not  contain a larger local
subalgebra than just the Lorentz group.  In reference [1] it was proposed
that one should replace
$G_{11}$ by a Kac-Moody algebra.   It was shown that the smallest such
algebra which contained $G_{11}$ was
$E_{11}$ and it was proposed to use the non-linear realisation of this
algebra with the local subalgebra  being the  Cartan involution $I_c$ 
invariant subalgebra,  denoted $I_c(E_{11})$. Indeed, the lowest level
generators of $E_{11}$  were just those of $G_{11}$. It followed from the
the work of reference [15] that  the non-linear realisation of $E_{11}$
at low levels contained  the bosonic fields of  eleven dimensional
supergravity and  the result was the bosonic sector of eleven dimensional
supergravity theory, provided one introduced space-time by putting  the
space-time translations into the group element  and took  the simultaneous
non-linear realisation with the conformal group.  However,  the  full
non-linear realisation of $E_{11}$ leads to a theory with 
infinite number of fields [1] in addition  to those of the 
bosonic sector of  eleven dimensional supergravity.  
\par
As is apparent from above, the  work of reference  [1]  contained two
open    questions. The first of which  is how can one 
 introduce space-time, or equivalently, include   the space-time
translations in a way that is consistent  with $E_{11}$? 
While, the second question concerns the conformal group. When used 
simultaneously  with the non-linear realisation of
$E_{11}$  the conformal group has the effect of turning
all the rigid transformations, at least at low levels,  of $E_{11}$ into
local transformations. For example, $GL(D)$ transformations become general
coordinate transformations [16] and the transformations of the three form
and six form fields  become local gauge transformations [15].
However, what was not answered was what algebra should replace the
conformal algebra when dealing with the  full
$E_{11}$ theory and furthermore  what was the closure of this latter 
algebra and the $E_{11}$ algebra? 
\par
An answer to the first question was
given in reference [17]; it was proposed to included  the space-time
translations by working with a representation of
$E_{11}$ that included them. This was the fundamental representation $l_1$
associated with the node on the far end of the longest tail of the
$E_{11}$ Dynkin diagram, usually called node one, which at low levels has
the generators [17] 
$$
P_a, Z^{ab}, Z^{a_1\ldots a_5}, Z^{a_1\ldots a_7,b}, Z^{a_1\ldots a_8},
Z^{b_1 b_2 b_3,a_1 ...a_8}, Z^{ (c d), a_1 ...a_9}, 
\ldots  
\eqno(1)$$
Hence, if one takes
the non-linear realisation of
$E_{11}\otimes_s l_1$, simultaneously with the conformal group,  one
finds at low levels the bosonic sector of eleven dimensional supergravity
theory. At low levels here  means up to and including the rank six field
of the $E_{11}$ part, but only the space-time translations in the $l_1$
part. The symbol $\otimes_s $ means semi-direct product. 
\par
The $l_1$ multiplet  contains generators with the same index
structure as the central charges of the eleven dimensional supersymmetry
algebra and indeed they can be identified with them [17]. This was not 
expected  as the $E_{11}$ computation just concerned the bosonic sector
and did not incorporate fermions and so also no supersymmetry. In fact
the $l_1$ representation  is  the multiplet of  brane charges; this is
 apparent at low levels, but is also true at all levels and it
provides an eleven dimensional origin for the exotic brane charges found
in string theories in lower spacetime dimensions [26]. 
\par
In the non-linear realisation of $E_{11}\otimes_s l_1$ one should
introduce a generalised space-time with one coordinate for each
generator in the $l_1$ representation, that is 
$$x^a, x_{ab},x_{a_1\ldots a_5}, x_{a_1\ldots a_7,b}, x_{a_1\ldots
a_8},x_{b_1 b_2 b_3,a_1 ...a_8}, x_{ (c d), a_1 ...a_9},
 \ldots 
\eqno(2)$$
If we restrict ourselves to level zero then $E_{11}$ becomes 
$GL(D)$ and the $l_1$ representation just contains   $P_a$ with the 
corresponding coordinate being the usual space-time  $x^a$. Using the
$l_1$ representation  has the natural interpretation; it
allows  all possible ways of measuring "space-time", in particular it 
uses all possible brane probes and not just the point particles
associated with the usual spacetime coordinate $x^a$. However, as the
$l_1$ representation has an  infinite number of generators so we have a
generalised space-time with an infinite number of coordinates. This  is
perhaps not too troubling as the non-linear realisation  is  an
effective theory, which hopefully encodes some of the most important
symmetries of  the underlying fundamental theory.  Nonetheless, to
build a non-linear realisation from an infinite number of coordinates
would seem an intimidating proposition, but it is the purpose of this
paper to make a first attempt by carrying out this calculation at low
levels. 
\par
It is worth noting that the $E_{11}\otimes_s l_1$ non-linear realisation
provides a natural setting for discussions of the generalised geometry
envisaged by Hitchin and others in that the generalised spacetime it
leads to admits a natural action of U-duality transformations. Indeed,
the precise extensions of space-time required to realise the known
duality symmetries  for M theory  dimensionally reduced  to
three and above dimensions was given in reference [20] on page 14. 
\par
The Cartan involution $I_c$ acts on the Chevalley generators by
$I_c(E_a)=-F_a$, $I_c(F_a)=-E_a$ and $I_c(H_a)=-H_a$. As such, it takes
a generator corresponding to a  root $\alpha$ to a generator with
the   root $-\alpha$. The  low level Cartan involution invariant
subalgebra of
$E_{11}$,  
$I_c(E_{11})$, which   was computed in reference [17].  and found to
contain the generators 
$$
J_{ab}, S_{a_1a_2a_3},S_{a_1\ldots a_6}, S_{a_1\ldots a_8,b},\ldots
\eqno(3)$$
Here $J_{ab}$ are just the usual Lorentz generators. The commutators of
 the generators of equation (3) follow from the underlying $E_{11}$
algebra. Those for the first three  generators were   found to be given by
replacing the generators by eleven dimensional $\gamma$-matrices of the
same rank.  Thus up to this level the commutators are consistent  those of
SL(32) which is just the algebra of 32 by 32 matrices with determinant
one for which a suitable basis are the eleven dimensional 
$\gamma$-matrices. The commutator
$[S_{a_1a_2a_3},S_{a_1\ldots a_6}]$ is also consistent with SL(32)
provided one sets 
$S_{b_1\ldots b_6}{}^{[a_1a_2,a_3]}={1\over 2}\delta_{[b_1}^{[a_3}
\epsilon^{a_1a_2]}{}_{b_1\ldots b_6] c_1\ldots c_4}
\gamma^{c_1\ldots c_4}$, see equation (5.1) of reference [17]. 
Hence, up to level three $I_c(E_{11})$ is  the same as $SL(32)$. 
We note that  $SL(32)$ is not a subalgebra, but a truncation of $E_{11}$
and above level three the two algebras are not the same. However, as we
will be interested in just the first three levels we can replace the local
subalgebra in the non-linear realisation, which was  
$I_c(E_{11})$,  by $SL(32)$. Up to level three this means keeping only the
generators $P_a$, $Z^{ab}$ and 
$Z^{a_1\ldots a_5}$ in the $l_1$ representation and so the coordinates 
$x^a, x_{ab},x_{a_1\ldots a_5}$.   
\par
It is an obvious consequence of reference [17]   that  
$\lambda_\alpha, \ \alpha=1,\ldots ,32$ is a representation of 
$I_c(E_{11})$ up to   level three as it is  a representation of
SL(32). Indeed, it is a representation of $I_c(E_{11})$ at all levels as
this algebra is generated by Lorentz transformations and
$S_{a_1a_2a_3}$,  both of which possess transformations on
$\lambda_\alpha$.  We can interpret $\lambda_\alpha$ as a spinor in
eleven dimensional space-time. As the first three generators of  this
$SL(32)$ coincide with those of
$I_c(E_{11})$ we can compute, from the underlying action of $E_{11}$, the
transformations they induce on the generators
$P_a$, $Z^{ab}$ and $Z^{a_1\ldots a_5}$ in the $l_1$ representation.
These agree with those induced from the   $SL(32)$  transformations of
the supercharges that occurs in the eleven dimensional supergravity
algebra. However, as the spacetime generators  $P_a$ that occur in the
$l_1$ representation of $E_{11}$  and in the supersymmetry algebra
are the same,   we can therefore identify  the  two central charges in the
supersymmetry algebra with the level one and level two   generators in
the $l_1$ representation [17]. 
\par
We now consider the role of the conformal group  in more
detail. Let us denote by  ${\bf conf}(E_{11}\otimes_s l_1)$ 
the algebra that should replace the
conformal algebra  used at low levels  when carrying out
the simultaneous 
 non-linear realisation of $E_{11}$ to all levels. We will 
denote this algebra   more generally, by 
${\bf conf}(G\otimes_s l_1)$ if we are considering a non-linear
realisation based on a Kac-Moody algebra $G\otimes_s l_1$ rather than
$E_{11}\otimes_s l_1$. 
\par
We note that ${\bf conf}(GL(D)\otimes_s l_1)= SO(2,D)$, as the
fundamental representation associated with node one is just a
vector of $SL(D)$  and so leads to the spacetime translations, i.e.
$l_1=T^D$. This algebra consists of 
$I_c(SL(D))=SO(1,D-1)$,  the translations $P_a$ and the special conformal
transformations $K^a$, both of which are  representations of $SO(1,D-1)$.
Clearly, 
${\bf conf}(E_{11}\otimes_s l_1)$  must contain
$SO(2,D)$ at level zero. Given that ${\bf conf}(E_{11}\otimes_s l_1)$ 
will be used in a simultaneous non-linear realisation with $E_{11}$ we
might expect it to contain the same local subalgebra, namely
$I_c(E_{11})$ as then one only has to take objects that transform under
this same local subalgebra   to ensure invariance under the closure of
both algebras. Up to level three this means it should contain the
generators 
$J_{ab}, S_{a_1a_2a_3},S_{a_1\ldots a_6}$,  that is in effect $SL(32)$,
Both $P_a$ and $K_a$ must belong to 
representations of $SL(32)$ up to level three. This is the
case for
$P_a$ as this is contained in $l_1$ which, up to level three, is the
second rank symmetric tensor representation $Z_{\alpha\beta}$ of 
$SL(32)$.  In fact
$K_a$ must also be contained in the representation
$\tilde Z^{\alpha\beta}$  and so must contain 
$\tilde K^a, \tilde Z_{ab},\tilde Z_{a_1\ldots a_5}$.  As such, up to
level three
${\bf conf}(E_{11}\otimes_s l_1)$ contains $GL(32)$, $Z_{\alpha\beta}$ and
$\tilde Z^{\alpha\beta}$.   An algebra that
contains all these generators in a natural way is $Sp(64)$,  indeed it is
the smallest  finite dimensional semi-simple algebra to do so.   The
commutators of $Sp(64)$  are given by
$$
[Z_{\alpha \delta},Z_{\gamma \beta}]=0,\ \ 
[Z_{\alpha \beta},R_{\gamma}{}^{ \delta}]=
-\delta_\alpha^\delta Z_{\gamma \beta}
-\delta_\beta^\delta Z_{\gamma \alpha}
\eqno(4)$$
and 
$$
[\tilde Z^{\alpha \delta},\tilde Z^{\gamma \beta}]=0,\ \ [ \tilde
Z^{\alpha
\beta},R^{\gamma}{}_{
\delta}]=
\delta_\gamma^\beta \tilde Z^{\alpha \delta}
+\delta^\alpha_\gamma \tilde Z^{\delta \beta}
\eqno(5)$$
as well as 
$$
[Z_{\alpha\beta},\tilde Z^{\gamma\delta}]= -\delta_\beta^\gamma
R_\alpha{}^\delta -\delta_\alpha^\delta
R_\beta{}^\gamma-\delta_\beta^\delta R_\alpha{}^\gamma -
\delta_\alpha^\gamma R_\beta{}^\delta
\eqno(6)$$
In these equation we may write 
$$
Z_{\alpha \beta}= P_a(\gamma^a C^{-1})_{\alpha\beta}+ 
Z^{a_1a_2}(\gamma_{a_1a_2} C^{-1})_{\alpha\beta}+
Z^{a_1\dots a_5}(\gamma_{a_1\ldots a_5} C^{-1})_{\alpha\beta},
\eqno(7)$$
and 
$$
\tilde Z^{\alpha \beta}= K^a(C\gamma_a )_{\alpha\beta}+ 
\tilde Z_{a_1a_2}(C\gamma^{a_1a_2} )_{\alpha\beta}+
\tilde Z_{a_1\dots a_5}(C\gamma^{a_1\ldots a_5} )_{\alpha\beta},
\eqno(8)$$
 We will return later  to
the question of what is  ${\bf conf}(E_{11}\otimes_s l_1)$ at all levels. 
\par
The use of the local subalgebra $Sp(64)$  is confirmed by
considering the supersymmetric extension discussed above. Now we must
have an algebra that contains the conformal group and the supersymmetry
algebra in eleven dimensions, but it is known [18] that the smallest such
algebra is
$Osp(1/64)$ which  contains as its Grassmann even part $Sp(64)$.  The
relations of 
$Osp(1/64)$ are those of equations (4) (5) and (6) and 
$$\{Q_\alpha, Q_\beta\}= Z_{\alpha \beta}, \ \  
[Q_\alpha,Z_{\gamma \delta}]=0,\  \ 
[Q_\alpha, R_\gamma{}^\delta]= -\delta_\alpha^\delta Q_\gamma,
\eqno(9)$$ 
$$
\{S^\alpha, S^\beta\}=\tilde Z^{\alpha \beta}, \ 
[S^\alpha,\tilde Z^{\gamma \delta}]=0,\  
[S^\gamma, R_\alpha{}^\beta]= \delta_\alpha^\gamma S^\beta,\ 
\{Q_\alpha, S^\beta\}= R_{\alpha}{}^ \beta, \ 
\eqno(10)$$
\par
It is straightforward  to find a supersymmetric generalisation of the
$l_1$ representation up to level three. One expects that the fermions will
only transform under the local subalgebra. This is indeed the case for
the spinors that occur in the dimensional reduction of the maximal
supergravity theories and for the inclusion  of spinors into the
$E_{10}$ theory [19]. Since up to level three the local subalgebra
coincides with SL(32) we may take as our Grassmann odd partners 
$\theta_\alpha,\ \alpha=1,\ldots , 32$. Hence, up to level three,  we have
the coordinates 
$$
x^{\alpha\beta} = x^a(C\gamma_a )^{\alpha\beta}+ 
x^{a_1a_2}(C\gamma_{a_1a_2} )^{\alpha\beta}+
x^{a_1\dots a_5}(C\gamma_{a_1\ldots a_5} )^{\alpha\beta},\ \ {\rm and }\
\ \theta_\alpha
\eqno(11)$$
The supersymmetry transformations are as usual given by 
$$
\delta x^{\alpha\beta}= i(\epsilon^\alpha\theta^\beta+\epsilon
^\beta\theta^\alpha),\ \ \delta\theta^\alpha=\epsilon^\alpha
\eqno(12)$$
The supercharge also transforms under SL(32) as in from equation
(9). 
\par
We could     carry out the computation
of the non-linear realisation of $E_{11}\otimes_s l_1$ up to level three,
in which case the fields are  the coefficients of the generators of
the $E_{11}$ part of the group element and they would be  functions of 
the first three coordinates of equation (2), while the coefficients of the
generators of the $l_1$ representation  in the group element are the
coordinates of the generalised spacetime in equation (2). However, it was
explained in [20] that brane dynamics in the presence of a non-trivial
background should also have an
$E_{11}$ symmetry and it will be much simpler to construct the analogous 
 point particle action up the   level three. The quantisation of
this system leads to  fields that depend on the three coordinates and so
  dynamics of the  kind first mentioned. 
\par
Subsequent to reference [1], but before reference [17],  it
was proposed [21] to consider a theory based only on the $E_{10}$
subalgebra of $E_{11}$, but a novel way was given of incorporating
space-time that was different, even at lowest levels, from that considered
in [1].  It was conjectured that  space was  contained in
$E_{10}$, in particular, the fields were taken to depend only on time and
their spatial derivatives were to be contained at higher levels in the
$E_{10}$ algebra. 
\par
Before discussing the role of $E_{11}$ in brane dynamics it will be
instructive to recall  the simpler dynamics of the bosonic particle
(0-brane) moving in $D$ dimensional Minkowski spacetime. 
 The generic point particle
possesses the symmetry  $ISO(1,D)$ which is non-linearly realised with
local  subgroup  $SO(D-1)$ which is linearly realised.  Thus one finds
Goldstone fields for the spacetime  translations and for the broken
Lorentz group.  Traditionally,  one does not include the space-time
translations,  or the supercharges in the supersymmetric case,  in the
local subgroup,  but it  has been shown  recently one can also include
them and as a result worldline  reparemeterisations and 
$\kappa$-symmetry naturally emerge from the non-linear realisation [27]. 
The massless bosonic particle possess more   symmetry, namely 
the conformal group 
$SO(2,D)$. In fact one can formulate the massless point particle action as
a non-linear realisation based on the conformal group
$SO(2,D)$ for an appropriate subgroup  and  depending
on the parameterisation of the coset one can find different
equivalent formulations some of which contain twistor variables in a
manifest way [22]. 
\par
It is also possible to construct bosonic branes coupled to gravity from 
the viewpoint of non-linear realisations [15]. We recall that gravity by
itself is just the non-linear realisation of $IGL(D)=GL(D)\otimes_s T^D$
when carried out simultaneously with the non-linear realisation of the
conformal group,  the local, subgroup for both groups being the
group
$SO(1,D-1)$ [16, 15]. In this formulation the fields associated to the
generators of spacetime  translations are the coordinates of space-time
and the other Goldstone fields, that is the graviton field, depend on
these space-time coordinates. To construct the dynamics of  the brane
coupled to gravity one  takes the same  symmetry group, i.e.
$IGL(D)$, but  the local subgroup is now 
$SO(1,p)\otimes SO(D-p-1)$. One also    lets  the "fields" associated
with the spacetime translations  and the broken Lorentz generators depend
on the  coordinates that parameterise  the brane world volume, but the
remaining Goldstone fields, that is the the graviton, depend on the latter
fields. We note that for a massless particle we should also have the
enlarged conformal symmetry.  This is consistent with the construction of
gravity from the simultaneous non-linear realisation of  
$IGL(D)$ and this group.  
\par
We now wish to consider the analogue of this discussion for a point 
particle of the formulation of M theory based on a non-linear
realisation of $E_{11}\otimes_s l_1$.   As explained above, the 
non-linear realisation in the absence of a brane is  based on 
$E_{11}\otimes_s l_1$ with  a local subgroup which is the Cartan
involution invariant subgroup $I_c(E_{11})\otimes_s l_1$. The 
algebra $I_c(E_{11})$ is  preserved by the vacuum when all the $E_{11}$ 
Goldstone fields are absent, that is we have no background and so it is
the analogue of
$SO(1,D-1)$ for the point particle. Thus the point
particle of M theory in the absence of background fields  should have as
a symmetry algebra $I_c(E_{11})\otimes_s
l_1$ and if it is massless it should  be invariant under the analogue
of the conformal group, that is 
${\bf conf}(E_{11}\otimes_s l_1)$. As pointed out above,  up to level
three  $I_c(E_{11})$ is just $SL(32)$ which acts linearly on
the coordinates $x^a, x_{ab},x_{a_1\ldots a_5}$ while  ${\bf
conf}(E_{11}\otimes_s l_1)$ is $Sp(64)$. As a result,  we require a point
particle action with these symmetries. For the supersymmetric case the
coordinates are those of equation (11) and the analogue of the conformal
group is
$Osp(1/64)$.  An action invariant under these symmetries is given by 
$$
\int d\tau {  \Pi^{\alpha\beta}} \lambda_\alpha\lambda_\beta
\eqno(13)$$
In this equation $\Pi^{\alpha\beta}=\dot
x^{\alpha\beta}-i(\theta^\alpha\dot \theta^\beta+\theta^\beta\dot
\theta^\alpha),
\alpha,\beta=1,\ldots , 32$ and  $x^{\alpha\beta}$ is the most general
symmetric matrix of equation (11), $\dot x^{\alpha\beta}={d
x^{\alpha\beta}\over d\tau}$ and $\lambda_\alpha$ and $\theta_\alpha$
are Majorana spinors. To obtain the non-supersymmetric case we just set
$\theta_\alpha=0$. The spinor
$\lambda_\alpha$ arises in taking the non-linear realisation of
$Osp(1/64)$, or in the non-supersymmetric case
$Sp(64)$, with an appropriate local subalgebra and it would be
interesting to  recover the above action from this perspective in a
similar way to how actions of this generic form  in lower
dimensions have been found by taking smaller algebras in the non-linear
realisation [22]. 
\par
 The action of equation (13) is obviously invariant under the
supersymmetry transformations of equation (12) and has a manifest  
linearly realised $SL(32)$ symmetry. It also has a non-linearly realised
$Osp(1/64)$ symmetry as required. 
\par
Actions of the  generic form of equation (13) have appeared [8-14] in the
literature  in the context of higher spin theories. Indeed,  upon
quantising [8-14] such an action one finds the so called unfolded
description [4-7] of the higher spin equations which have, in addition to 
the usual coordinates $x^a$ of spacetime, a number of bosonic coordinates
which generically lead to an infinite number of higher spin fields. A
number of bosonic  and the
supersymmetric actions have been quantised  and the towers of resulting
higher spins found [8-14]. 
\par
However, the action of equation (13) has not been quantised for the
eleven dimensional case, nor  have the corresponding
actions for the IIA and IIB cases. 
The action for the  IIA case contains a Majorana spinor of  the
ten dimensional theory and the Grassmann even coordinates are just those
corresponding to the central charges of the supersymmetry algebra that
occur for the IIA theory, i.e. $x^a, x_{ab}, x_{a_1\ldots a_5},
x_{a_1\ldots a_6}, x_{a_1\ldots a_9}, x_{a_1\ldots a_{10}}$.  However,
for the IIB theory we take the spinors to be two Majorana-Weyl spinors of
the same chirality and the Grassmann even coordinates are given by 
$$
x^{\alpha i\beta j}=( C\gamma_a P_\pm)^{\alpha \beta}
\delta^{ij}x^a
+
( C\gamma^{a} P_\pm)^{\alpha \beta}
x_a^{ij}
+
$$
$$
(C\gamma^{a_1 a_2a_3} P_\pm )^{\alpha \beta}
\epsilon^{ij} x_{a_1 a_2 a_3}
+
( C\gamma^{a_1\dots a_5} P_\pm)^{\alpha \beta}
x_{a_1\dots a_5}^{ ij}, \ \ i,j=1,2
\eqno(14)$$
where $x_a^{ij}$ is symmetric and traceless in its $ij$ indices, 
$x_{a_1\dots a_5}^{ ij}$ is only symmetric in its $ij$ indices, but 
self-dual, or anti-self-dual, in its $a_1\dots a_5$ indices. 
\par
If we consider the action of equation (13) for $x^{ab}=0=x^{a_1\ldots
a_5}$, and  impose the constraint  $(\bar\lambda\gamma^a\lambda
)^2=0$, then it is equivalent to the Brink-Schwarz action [25] whose
quantisation is known to lead to the massless states of eleven 
dimensional supergravity. The actions for the IIA and IIB cases with all
the coordinates except $x^a$ set to zero are equivalent to the
corresponding  Brink-Schwarz actions. On the other hand, if we set
$\theta_\alpha=0$,  its quantisation leads to an infinite number of
higher spin fields. Thus the quantisation of the action of equation (13)
should lead to an infinite number of propagating higher spin fields which
occur in multiplets of spacetime supersymmetry and have as their lowest
states those of eleven dimensional supergravity. It would be interesting
to find out precisely what higher spin fields it propagates. Similar
conclusions hold for the IIA and IIB cases and one cannot help wondering
if one then finds some connection with the states of the IIA and IIB
string theories. 
\par
In this paper we have constructed, at low levels,   the dynamics of a 
point particle associated with a formulation  of M theory based on 
  a  non-linear realisation of $E_{11}\otimes_s l_1$. We have
isolated the symmetries that such a point particle should possess at low
levels, and we have written down a corresponding action. It is of the
generic form of the point particle actions whose quantisation in less
dimensions and for few supersymmetries have been previously considered
and have been shown to lead to an infinite set of higher spin fields. The
final quantised form of equation (13) will be a dynamics involving
fields which are functions of the infinite set of coordinates
associated with the $l_1$ representation. It would be interesting to
see if this was the same as  the non-linear realisation of
$E_{11}\otimes_s l_1$ with local subalgebra $I_c(E_{11})$, however, in
both cases one   expects to find  an infinite
tower of higher spin fields.   It is possible that one finds in this way
the string states contained in the non-linear realisation. We note  that
the dependence of the fields on the additional bosonic coordinates is
responsible for the propagating the higher spin states.  Very recently
[23], it was shown that the fields in the adjoint representation  contain
all possible dual formulations of the on-shell degrees of freedom of 
eleven dimensional supergravity and that some other of these  fields are
likely to be responsible for the gauged supergravities in lower
dimensions. Of course it is not excluded that the fields in the adjoint
representation can also contribute to higher propagating degrees of
freedom, however, a possible  picture is that the fields associated with
the adjoint representation of 
$E_{11}$, apart from those at levels zero and one, make the symmetry
manifest and that the dependence on the additional coordinates lead to
higher spin propagating states.  It is important to stress that we have
been working only up to level three and so with  a truncation of  the real
theory. In particular,  the underlying non-linear realisation containing
the
$l_1$ representation possess an infinite number of coordinates and so 
the higher order results will be rather different. 
\par
We close this paper by returning to the discussion of  the
   algebra  ${\bf conf}(E_{11}\otimes_s l_1)$. As discussed above 
this algebra should contain $I_c(E_{11})$ and the conformal algebra. 
The generators $P_a$ and $K^a$ must separately belong to representations
of  $I_c(E_{11})$. We recall that $P_a$ is the
lowest component of the $l_1$ representation of $E_{11}$ and it is
natural to take $K^a$ to belong to  the representation that is obtained
from the $l_1$ representation by the Cartan involution. This latter
representation,  is just the one with lowest weight
$-\Lambda_1$, where $\Lambda_1$ is the highest weight of the
fundamental representation associated with node one. It is obtained from
the
$l_1$ representation by acting with the Cartan involution $I_c$ and so we
may write it as $I_c(l_1)$. Assuming this we now give a plausible
construction of
${\bf conf}(E_{11}\otimes_s l_1)$.  The
$l_1$ representation can be constructed by adding one node to the Dynkin
diagram of $E_{11}=E_8^{+++}$ attached by a single line to node one.
That is consider  the algebra $E_8^{++++}$;  the $l_1$ representation is
then just the generators of level one with respect to this new node [24]. 
The  generator $K^a$ belongs to the representation of $E_8^{+++}$
which is just all generators of the algebra $E_8^{++++}$ with level $-1$
with respect to the additional node. We can now construct the algebra 
${\bf conf}(E_{11}\otimes_s l_1)$ from the algebra $E_8^{++++}$. We 
assume that the commutator of two generators from the $l_1$
representation vanish and adopt  the same relation for any two generators
of the $I_c(l_1)$ representation. This is tantamount to saying we only
want  generators of level $0$ and
$\pm 1$ in  ${\bf conf}(E_{11}\otimes_s l_1)$. 
Given a generator $A$ in the $l_1$ representation, we can  form the
combinations $R=A+I_c(A)$ and  $S=A-I_c(A)$. Clearly, 
$I_c(R)=R$ and $I_c(S)=-S$ and so the commutators $[R_1, R_2]$,  and
$[S_1, S_2]$ are invariant under $I_c$
and are of level zero with respect to the additional node. As such, they
are in
$E_8^{+++}$ and so  belong to $I_c(E_{11})$ and we
take their values in ${\bf conf}(E_{11}\otimes_s l_1)$ to be just those
that arise in the  $E_8^{++++}$ algebra. 
However, the commutator $[R_1, S_2]$ is of level zero with respect to
the new node and so it is in  $E_8^{+++}$, but it is odd under
$I_c$. We now  truncate the terms that appear on the right-hand side of
this commutator. The simplest possibility to take all such terms on the
right-hand side to be zero, except 
$D=-\sum _aK^a{}_a$, which we keep. One could also envisage
a less restrictive condition, however, this would have to be compatible
with the Jacobi identities, in particular those involving $I_c(E_{11})$, 
and the level analysis associated within $E_{11}$ itself. Thus we have
specified an algebra for the set of generators which are those in  
$I_c(E_{11})$, $l_1$, $I_c(l_1)$ and $D$ which we suggest to be 
${\bf conf}(E_{11}\otimes_s l_1)$.  Should we apply this method  to
$GL(D)\otimes_s l_1$ it does lead to the conformal group $SO(2,D)$ and 
up to the first three levels ${\bf conf}(E_{11}\otimes_s l_1)$ is
$Sp(64)$.  This construction also works if we replace
$E_8$ by any finite dimensional semi-simple Lie algebra $G$ to construct 
${\bf conf}(G^{+++}\otimes_s l_1)$. This latter algebra  is
relevant for  theories  based on the non-linear realisation of
$G^{+++}\otimes _s l_1$ with which one can form the simultaneous
non-linear realisation. Such theories include the extensions of pure
gravity and the effective action of the closed bosonic string [1,28]. 
\medskip
{\bf Acknowledgments}
\medskip
I would like to thank  the hospitality  of the physics department of
the University of Canterbury at  Christchurch, New Zealand. I
would    like to thank  Andrew Pressley for discussions on group
theory, also  discussions with Misha Vasiliev and Igor Bandos,   the  
support of a PPARC senior fellowship PPA/Y/S/2002/001/44 and  
of    a  PPARC rolling grant   PP/C5071745/1 and  the EU Marie Curie,
research training network grant HPRN-CT-2000-00122. 

\medskip
{\bf References}
\medskip
\item{[1]} P. West, {\it $E_{11}$ and M Theory}, Class. Quant.
Grav. {\bf 18 } (2001) 4443,  hep-th/010408.
\item{[2]} C. Fronsdal, {\it Massless Particles, Ortosymplectic Symmetry
and Another Type of Kaluza--Klein Theory}, Preprint UCLA/85/TEP/10, in
``Essays on Supersymmetry", Reidel, 1986 (Mathematical Physics
Studies, v. 8).
\item{[3]}  M. A. Vasiliev,
 {\it Unfolded representation for relativistic equations in (2+1) anti-De
Sitter space}, 
Class. Quant. Grav.  {\bf 11} (1994) 649 ;
\item{[4]} M.A. Vasiliev, {\it Relativity, Causality, Locality,
Quantization and Duality in the $Sp(2M)$ Invariant Generalized
Space-Time}, in {\it Multiple facets of quantization and supersymmetry.
Marinov's Memorial Volume} (eds. M.Olshanetsky and A.Vainshtein) pp.
826-872, hep-th/0111119.
\item{[5]}  M.A. Vasiliev, {\it  Conformal Higher Spin Symmetries of 4d
Massless Supermultiplets and $osp(L,2M)$ Invariant Equations in
Generalized  (Super)Space},
Phys. Rev. {\bf D66} (2002) 066006,  hep-th/0106149.
\item{[6]}  M. A. Vasiliev, {\it Progress in higher spin gauge theories},
hep-th/0104246,
{\it Higher-Spin Theories and $Sp(2M)$ Invariant Space--Time},
 hep-th/0301235.
\item{[7]} M.A.Vasiliev, {\it  Actions, Charges and Off-Shell Fields in
the Unfolded Dynamics Approach},  Int.J.Geom.Meth.Mod.Phys. 3 (2006)
37-80, hep-th/0504090.Ê
\item{[8]} I. Bandos and J. Lukierski, {\it Tensorial Central Charges and
New Superparticle Models with Fundamental Spinor Coordinates}, Mod. Phys.
Lett. {\bf 14} (1999) 1257, hep-th/9811022.
\item{[9]} I. Bandos and J. Lukierski, {\it New Superparticle Models
Outside the HLS Supersymmetry Scheme}, Lect. Not. Phys. {\bf 539} (2000)
195 (hep-th/9812074).
\item{[10]} {{I. Bandos}}, J. Lukierski and D. Sorokin,
{\sl Superparticle Models with Tensorial Central Charges},
Phys. Rev. {\bf D61} (2000) 045002,    hep-th/9904109.
\item{[11]} I.~A.~Bandos, J.~Lukierski, C.~Preitschopf and D.~P.~Sorokin,
{\it OSp supergroup manifolds, superparticles and supertwistors},
Phys.\ Rev.\ D {\bf 61} (2000) 065009, hep-th/9907113.
\item{[12]}  V.E. Didenko, M.A. Vasiliev, J.Math.Phys.45:197-215,2004, 
hep-th/0301054; O.A. Gelfond,  M.A.
Vasiliev, Theor.Math.Phys.145:1400-1424,2005, Teor.Mat.Fiz.145:35-65,2005,
hep-th/0304020.   
\item{[13]} M. Plyushchay, D. Sorokin and M.
Tsulaia,{\it Higher spins from tensorial charges and $OSp(N|2n)$
symmetry}, JHEP 0304 (2003) 013, hep-th/0301067; 
  M. Plyushchay, D. Sorokin and M. Tsulaia, {\it GL
Flatness of $OSp(1|2n)$ and Higher Spin Field Theory from Dynamics in
Tensorial Spaces}, hep-th/0310297; 
\item{[14]} I.  Bandos, P. Pasti, D. Sorokin, M. Tonin, {\it Superfield
Theories in Tensorial Superspaces and the Dynamics of Higher Spin Fields},
JHEP 0411 (2004) 023, hep-th/0407180; 
 I.  Bandos, X. Bekaert, J. de Azcarraga, D. Sorokin, M. Tsulaia, {\it
Dynamics of Higher Spin Fields and Tensorial Space}, hep-th/0501113. 
  Ê\item{[15]} P.~C. West, {\it Hidden superconformal symmetry in {M}
   theory },  JHEP {\bf 08} (2000) 007,  hep-th/0005270.
\item{[16]} V. Ogievetsky, Lett. {\it Infinite-dimensional algebra of
general  covariance group as the closure of the finite dimensional
algebras  of conformal and linear groups}, Nuovo. Cimento, 8 (1973) 988; 
 A. Borisov and V. Ogievetsky,  {\it Theory of dynamical affine and
confomral  symmetries as the theory of the gravitational field}, 
Teor. Mat. Fiz. 21 (1974) 329. 
\item{[17]} P. West, {\it $E_{11}$, SL(32) and Central Charges},
Phys. Lett. {\bf B 575} (2003) 333-342, {\tt hep-th/0307098}
\item{[18]} J. van Holten and A. van Proeyen,{\it  $N=1$ supersymmetry
algebras  in $d=2,3,4$ mod 8}, J. Phys A, 15 (1982) 376. 
\item{[19]} T. Damour, A. Kleinschmidt, H. Nicolai, {\it Hidden
symmetries and the fermionic sector of eleven-dimensional supergravity}, 
 Phys.Lett. B634 (2006) 319-324, hep-th/0512163;  Ê
S. de Buyl, M. Henneaux and  L. Paulot, {\it  Extended E8 Invariance of
11-Dimensional Supergravity}, JHEP 0602 (2006) 056, hep-th/0512292. 
\item{[20]} P. West, {\it Brane dynamics, central charges and
$E_{11}$},  hep-th/0412336. 
\item{[21]}  T. Damour, M. Henneaux and H. Nicolai, {\sl $E_{10}$ and a
``small tension expansion'' of M-theory}, Phys. Rev. Lett. {\bf 89}
(2002) 221601, {\tt hep-th/0207267}
\item{[22]} P. Howe and P. West, {\it The Conformal Group, Point
Particles and Twistors}, Int. Jour. Mod. Phys. A7 (1992) 6639. 
\item{[23]} F. Riccioni and  P.  West, {\it  Dual fields and $E_{11}$},
hep-th/0612001. 
\item{[24]}  A. Kleinschmidt and P. West, {\it  Representations of G+++ and
the role of space-time},  JHEP 0402 (2004) 033,  hep-th/0312247.
\item{[25]} L. Brink and J. Schwarz, Phys. Lett. 100B (1981) 310. 
\item{[26]} P. West,  {\it $E_{11}$ origin of Brane charges and U-duality
multiplets}, JHEP 0408 (2004) 052, hep-th/0406150. 
\item{[27]}  J. Gomis, K. Kamimura and  Peter West, {\it 
Diffeomorphism, kappa transformations and the theory of non-linear
realisations }, JHEP 0610 (2006) 015, hep-th/0607104. 
\item{[28]} N. Lambert and P. West, {\it Coset
Symmetries in Dimensionally Reduced Bosonic String Theory}, Nucl. Phys.
{\bf B 615} (2001) 117,hep-th/0107209;
 F. Englert, L. Houart, A. Taormina and P. West, JHEP {\bf 0309} (2003)
020 {\it The Symmetry of M-Theories}, hep-th/0304206.

\end